\documentclass[prd,nofootinbib,twocolumn, floatfix, superscriptaddress]{revtex4}
\usepackage[utf8]{inputenc}
\pdfoutput=1
 
\usepackage{graphicx}
\usepackage{epsfig}
\usepackage{bm}
\usepackage{amssymb}
\usepackage{float}
\usepackage{amsmath}
\usepackage{dcolumn}
\usepackage{cancel}
\usepackage[colorlinks]{hyperref}
\usepackage[usenames, dvipsnames]{color}

\hypersetup{
     breaklinks=true, 
    pdfstartview={FitH},  
    colorlinks=true, 
    linkcolor=blue,  
    citecolor=red,  
    filecolor=magenta,  
    urlcolor=blue, 
    anchorcolor=green,  
    linktocpage=true
}

\providecommand{\U}[1]{\protect\rule{.1in}{.1in}}

\newcommand{\be}{\begin{equation}}
\newcommand{\ee}{\end{equation}}

\newcommand{\mincir}{\raise
-3.truept\hbox{\rlap{\hbox{$\sim$}}\raise4.truept\hbox{$<$}\ }}
\newcommand{\magcir}{\raise
-3.truept\hbox{\rlap{\hbox{$\sim$}}\raise4.truept\hbox{$>$}\ }}

\ifx\pdfoutput\relax\let\pdfoutput=\undefined\fi
\newcount\msipdfoutput
\ifx\pdfoutput\undefined\else
\ifcase\pdfoutput\else
\ifx\paperwidth\undefined\else
\ifdim\paperheight=0pt\relax\else\pdfpageheight\paperheight\fi
\ifdim\paperwidth=0pt\relax\else\pdfpagewidth\paperwidth\fi
\fi\fi\fi

\hypersetup{colorlinks=true,
	breaklinks=true,
	pdfstartview=Fit,
	linkcolor=blue,
	citecolor=blue,
	urlcolor=blue}

\begin{document}
\title{Constraints on Barrow and Tsallis  Holographic Dark Energy from DESI 
DR2 BAO data}

\author{Giuseppe Gaetano Luciano}
\email{giuseppegaetano.luciano@udl.cat}
\affiliation{Departamento de Qu\'{\i}mica, F\'{\i}sica y Ciencias Ambientales y 
del Suelo,  Escuela Polit\'ecnica Superior -- Lleida, Universidad de Lleida, Av. 
Jaume II, 69, 25001 Lleida, Spain}

\author{Andronikos Paliathanasis}
\email{anpaliat@phys.uoa.gr}
\affiliation{School for Data Science and Computational Thinking, Stellenbosch 
University,
44 Banghoek Rd, Stellenbosch 7600, South Africa}
\affiliation{Departamento de Matem\`{a}ticas, Universidad Cat\`{o}lica del 
Norte, Avda.
Angamos 0610, Casilla 1280 Antofagasta, Chile}
\affiliation{National Institute for Theoretical and Computational Sciences (NITheCS), South Africa}

\author{Emmanuel N. Saridakis}
\email{msaridak@noa.gr}
\affiliation{National Observatory of Athens, Lofos Nymfon 11852, Greece}
\affiliation{CAS Key Laboratory for Research in Galaxies and Cosmology, School 
of Astronomy and Space Science, University of Science and Technology of China, 
Hefei 230026, China}
\affiliation{Departamento de Matem\`{a}ticas, Universidad Cat\`{o}lica del 
Norte, Avda.
Angamos 0610, Casilla 1280 Antofagasta, Chile}

\begin{abstract} 
Barrow and Tsallis  Holographic Dark Energy (HDE) are two recently proposed 
extensions of the standard HDE framework, incorporating generalized corrections 
to horizon entropy through the use of Barrow and Tsallis  entropies. Tsallis 
entropy arises from non-extensive statistical phenomena  which 
account for long-range correlations and deviations from additivity, while 
Barrow entropy emerges from  quantum-gravitational effects  on the horizon 
geometry, associated with fractal modifications and deformations. At the 
cosmological level, both scenarios lead to the same equations, nevertheless 
the involved parameters obey different theoretical bounds.    In 
this work, we use observational data from Supernova Type Ia (SNIa), Cosmic 
Chronometers (CC) and Baryonic acoustic oscillations (BAO), including the 
recently released DESI DR2 dataset, to place constraints on both scenaria. We 
show that both can be in agreement with observations, although they cannot 
alleviate the $H_0$ tension. However, 
applying information criteria we deduce that both of them are not favoured 
comparing to $\Lambda$CDM concordance cosmological paradigm.

\end{abstract}


\maketitle

\section{Introduction}
\label{Intro}
Holographic dark energy  (HDE) offers an alternative theoretical approach to 
the dark energy problem, based on the holographic principle 
\cite{tHooft:1993dmi,Susskind:1994vu,Bousso:2002ju}. This framework arises from 
the proposed relationship between the ultraviolet (UV) cutoff and the maximum 
permissible infrared (IR) scale in effective quantum field theory 
\cite{Cohen:1998zx}. In this setting, the vacuum energy density is interpreted 
as a manifestation of dark energy at cosmological scales 
\cite{Li:2004rb,Wang:2016och}.

A basic point in implementing the  holographic principle in a cosmological 
context lies in identifying a physically consistent IR cutoff and understanding 
its associated thermodynamic implications. The standard consideration in this 
direction is that the entropy of a cosmological horizon scales with its surface 
area rather than its volume, similarly to the Bekenstein-Hawking entropy 
relation for black holes \cite{Bekenstein:1973ur,Hawking:1975vcx}. 
Building on this idea,  the original HDE model adopted the future event horizon 
as the IR cutoff and applied the Bekenstein-Hawking area law as the 
corresponding entropy bound, resulting to a scenario that 
  shows compatibility with a broad range of observational data 
\cite{Zhang:2005hs,Li:2009bn,Feng:2007wn,Zhang:2009un,Lu:2009iv}.

Nevertheless, one can extend the original scenario in many ways.
 Initial research efforts have investigated the use of alternative IR cutoffs 
and the possibility of interactions between the Universe dark components 
\cite{Wang:2016och, Wang:2016lxa}. More recently, driven by considerations from 
generalized statistical mechanics, several new HDE models have been developed 
based on modified entropy frameworks, including those proposed by Tsallis 
\cite{Tsallis:1987eu, Tsallis:2013}, Kaniadakis \cite{kaniadakis2001non}, Renyi 
\cite{renyi1961entropy} and Barrow \cite{Barrow:2020tzx}, among others, and 
their cosmological applications have been widely studied
\cite{Horvat:2004vn,Huang:2004ai,Pavon:2005yx,
Wang:2005jx,Nojiri:2005pu,Wang:2005ph,Setare:2006wh,Setare:2008pc,
Gong:2004fq,Saridakis:2007cy,Setare:2007we,Cai:2007us,Saridakis:2007ns, 
Jamil:2009sq,Micheletti:2009jy,Aviles:2011sfa,Chimento:2011pk,Pourhassan:2017cba,Tavayef:2018xwx,Saridakis:2018unr, 
Nojiri:2019kkp,Geng:2019shx,DAgostino:2019wko,Saridakis:2020zol,
Drepanou:2021jiv,Luciano:2022hhy,Luciano:2022ffn,Nakarachinda:2023jko,Mamon:2020spa,Ghaffari:2022skp,Luciano:2023roh,Luciano:2023wtx}.

Notably, the Tsallis and Barrow  entropies have drawn particular attention. 
Although they originate from fundamentally different physical considerations, 
both lead to a power-law deformation of the Bekenstein-Hawking entropy in the 
form
\begin{equation}
\label{PLdef}
    S_\delta={\left(\frac{A}{A_0}\right)}^\delta\,,
\end{equation}
where $A = L^2$  is the horizon area of the holographic system and $A_0=4L_p^2$ 
represents 
the Planck area.  The exponent \( \delta \) quantifies the deviation from the 
standard Bekenstein-Hawking area law and encodes the underlying effects specific 
to each framework. In   Tsallis formulation, it is related to 
to the degree of non-additivity characterizing the complex system under consideration. 
Specifically,  values of \( \delta < 1 \) correspond to a sub-additive scaling 
of the number of accessible microstates with respect to the horizon area, 
indicating that the total entropy grows more slowly than linearly with system 
size.  Conversely, \( \delta > 1 \) represents a super-additive regime, 
potentially signaling an overcounting of degrees of freedom or contributions 
from high-energy, nonlocal effects. The standard Bekenstein-Hawking entropy is 
recovered for \( \delta = 1 \). 

In contrast,  within the Barrow framework, the deformation of the 
Bekenstein-Hawking area law originates from quantum gravitational effects that 
induce a fractal-like structure on the black hole or cosmological horizon 
\cite{Barrow:2020tzx}. These quantum fluctuations are encoded through a 
dimensionless parameter \( \Delta \in [0,1] \), which quantifies the degree of 
such spacetime irregularity. This leads to a modified entropy-area relation 
$S_\Delta$ of the form \eqref{PLdef}, provided one replaces 
$\delta\rightarrow1+\Delta/2$.
In this scenario, \( \Delta = 0 \) recovers the classical Bekenstein-Hawking 
result,  while \( \Delta = 1 \) corresponds to a maximally deformed, highly 
quantum-corrected case. Although Barrow's original formulation assumes 
$\Delta>0$ to model sphereflake-like deformations of the horizon geometry, more 
general arguments from condensed matter systems \cite{Tang} and quantum field 
theory \cite{Kogut}   support the possibility of negative values, too. 
Moreover, in the regime of small deviations from the standard holographic 
scaling, the Tsallis–Barrow entropy reduces to a logarithmic correction to the 
area law, consistent with predictions from various quantum gravity approaches 
\cite{Banerjee:2011jp,Kaul:2000kf,Carlip:2000nv}.

The extension of  the HDE model using Tsallis and Barrow entropies has been 
proposed and investigated in \cite{Tavayef:2018xwx,Saridakis:2020zol}, 
demonstrating that this framework can successfully reproduce the thermal history 
of the Universe, including the sequence of matter- and dark energy-dominated 
eras. Interestingly,   the 
entropic exponent plays a significant role in determining the behavior of the 
dark energy equation of state (EoS), allowing for quintessence-like dynamics, 
entry into the phantom regime or even a crossing of the phantom divide during 
cosmic evolution \cite{Saridakis:2020zol}.

On the other hand, the recent data releases from DESI,  including the DESI DR2 
dataset, have already proven to be a valuable resource for placing stringent 
constraints on a wide spectrum of cosmological models~\cite{DESI:2025zgx}. 
Thanks to the exceptional precision of Baryon Acoustic Oscillation (BAO) 
measurements, this dataset has enabled detailed tests of numerous extensions to 
the standard \(\Lambda\)CDM model. In particular, it has been used to constrain 
dynamical dark energy 
frameworks~\cite{Ormondroyd:2025iaf,You:2025uon,Gu:2025xie,Santos:2025wiv,
Li:2025cxn,Alfano:2025gie,Carloni:2024zpl}, early dark energy 
scenarios~\cite{Chaussidon:2025npr} and a variety of scalar field theories with 
both minimal and non-minimal 
couplings~\cite{Anchordoqui:2025fgz,Ye:2025ulq,Wolf:2025jed}. Moreover, BAO data 
have been employed to investigate quantum-gravity-inspired approaches such as 
those derived from the Generalized Uncertainty 
Principle~\cite{Paliathanasis:2025dcr}, as well as interacting dark sector 
models~\cite{Shah:2025ayl,Silva:2025hxw,Pan:2025qwy}. Additional applications 
include astrophysical tests~\cite{Alfano:2024jqn}, model-independent 
cosmographic reconstructions~\cite{Luongo:2024fww}, a broad range of modified 
gravity/entropy 
theories~\cite{Yang:2025mws,Li:2025cxn,Paliathanasis:2025hjw,Tyagi:2025zov,
Luciano:2025hjn}, as well as various other models and scenarios
\cite{Brandenberger:2025hof,Paliathanasis:2025cuc,Ishiyama:2025bbd,
  Wang:2025ljj,
 Akrami:2025zlb,Colgain:2025nzf,
Plaza:2025gjf,  Toda:2025dzd,Dinda:2025svh,
  Kumar:2025etf,Mirpoorian:2025rfp,
deSouza:2025rhv,Scherer:2025esj,Preston:2025tyl,Abedin:2025yru,
Wang:2025vfb,
Bayat:2025xfr,Cai:2025mas,Ye:2025ark,Andriot:2025los,RoyChoudhury:2025dhe,
vanderWesthuizen:2025iam}.

In this work, we use  observational data from the recent DESI DR2 release to 
impose constraints on the Tsallis-Barrow HDE model. Our primary objective is to 
derive observational bounds on the deformation parameter, which quantifies the 
departure from the standard entropy-area relation. The plan of the work is as 
follows. In Section \ref{model} we present Tsallis and Barrow holographic dark 
energy scenaria. Then, in Section \ref{Obs} we use datasets from Supernova Type 
Ia (SNIa), Cosmic 
Chronometers (CC) and Baryonic acoustic oscillations (BAO) observations, 
including the 
recently released DESI DR2 data, in order to extract constraints on the model 
parameters. Finally, in Section \ref{Conc} we summarize the obtained results.

\section{Tsallis and Barrow Holographic Dark Energy}
\label{model}

Following  \cite{Tavayef:2018xwx,Saridakis:2020zol}, in this section we derive 
the generalized HDE model using  \eqref{PLdef}. For simplicity, we explicitly 
focus on the case of Barrow entropy, while noting that the Tsallis scenario can 
be straightforwardly recovered by applying the substitution \( \Delta 
\rightarrow 2(\delta - 1) \) (see the discussion in the Introduction).

In the standard  HDE description, the energy density is constrained by the 
inequality \( \rho_{\mathrm{DE}}\hspace{0.4mm} L^4 \leq S \), where \( L \) 
denotes the IR cutoff length scale. Assuming that the entropy scales with the 
horizon area as \( S \propto A \propto L^2 \)~\cite{Wang:2016och}, one recovers 
the conventional HDE model. However, replacing the Bekenstein-Hawking entropy 
with the Tsallis-Barrow-modified entropy \eqref{PLdef} leads to
\begin{equation}
\label{rde}
    \rho_{\mathrm{DE}}=C L^{\Delta-2}\,,
\end{equation}
where $C$ has dimensions $[L]^{-2-\Delta}$.
When  the deformation parameter \( \Delta \) vanishes, the expression naturally 
reduces to the standard HDE density, \( \rho_{\mathrm{DE}} = 3c^2 M_p^2 L^{-2} 
\), where \( M_p \) is the (reduced) Planck mass and \( c^2 \) is the 
dimensionless model parameter, with \( C = 3c^2 M_p^2 \). In contrast, when \( 
\Delta \neq 0 \), the  corrections introduced by the Tsallis-Barrow entropy 
become relevant, causing deviations from the conventional HDE form and resulting 
in modified cosmological dynamics.

In order to investigate  the implications of these non-standard evolutionary 
behaviors, we consider a spatially flat, homogeneous and isotropic 
Universe, described by the $(3\hspace{-0.2mm}+\hspace{-0.2mm}1)$-dimensional  
Friedmann–Robertson–Walker (FRW) metric
\begin{equation}
ds^2 =  h_{\mu\nu} \, dx^\mu dx^\nu \ +  \ \tilde{r}^2\left(d\theta^2 \ + \ 
\sin^2\theta\, d\phi^2\right)\, ,
\label{FRW}
\end{equation}
where the areal  radius is given by $\tilde{r} = a(t)\hspace{0.2mm}r$, and the 
coordinates are specified as $x^0 = t$ and $x^1 = r$, respectively. The metric 
tensor $h_{\mu\nu}$ describes the two-dimensional $(t, r)$ subspace and is 
expressed as $h_{\mu\nu} = \mathrm{diag}(-1, a^2)$, with $a(t)$ denoting the 
time-dependent scale factor.

Regarding  the choice of the IR cutoff \( L \) in HDE models, we here consider 
the most widely accepted definition in the literature, namely the future event 
horizon \cite{Li:2004rb} 
\begin{equation}
\label{IRc}
R_h = a \int_t^{\infty} \frac{dt'}{a(t')} = a \int_a^\infty \frac{da'}{Ha'^2} \, ,
\end{equation}
where $H=\dot a/a$ the Hubble parameter and the overdot denotes time 
derivative. Substituting  this into relation \eqref{rde}, the energy density of 
Barrow holographic dark energy (BHDE) takes the form
\begin{equation}
\label{BHDE}
    \rho_{\mathrm{DE}}=C R_h^{\Delta-2}\,.
\end{equation}

We additionally  assume that the Universe contains both the standard matter 
component, modeled as a perfect fluid of energy density $\rho_m$, and the 
previously introduced BHDE. Under this setup, the two Friedmann equations read
\begin{eqnarray}
\label{FFE}
\rho_m+\rho_{\mathrm{DE}}&=&3M_p^2H^2\,,\\[2mm]
\rho_m+p_m+\rho_{\mathrm{DE}}+p_{\mathrm{DE}}&=&-2M_p^2\dot H\,,
\label{SFE}
\end{eqnarray}
where $p_{\mathrm{DE}}$  and $p_m$ denote the pressure of BHDE and matter, 
respectively. By introducing the fractional densities
\begin{equation}
    \label{frde}
    \Omega_m\equiv \frac{\rho_m}{3M_p^2H^2}\,,\qquad  \Omega_{\mathrm{DE}}\equiv 
\frac{\rho_\mathrm{DE}}{3M_p^2H^2}\,,
\end{equation}
Eq. \eqref{FFE} can be equivalently expressed as 
\begin{equation}
\label{eqform}
\Omega_m+\Omega_{\mathrm{DE}}=1\,.
\end{equation}
Additional conditions arise from the continuity equation, which, for the matter 
and BHDE sectors,  is expressed as
\begin{eqnarray}
\label{contmat}
    \dot\rho_m + 3H\left(\rho_m + p_m\right) 
&=& 0\,,\\[2mm]
\dot\rho_{\mathrm{DE}} + 3H\left(\rho_{\mathrm{DE}} + p_{\mathrm{DE}}\right) 
&=& 0\,,
\label{contDE}
\end{eqnarray}
respectively.
In particular,  assuming that the matter component is pressureless dust (\( p_m 
= 0 \)), Eq. \eqref{contmat} determines the evolution of the matter energy 
density as $\rho_m=\rho_{m0}/a^3$,
where \( \rho_{m0} \) denotes the present-day matter  energy density, 
corresponding to \( a_0 = 1 \) (hereafter, a subscript ``0'' indicates the 
present value of a given quantity). 
Substituting this result into Eq. \eqref{frde} gives 
\begin{equation}
\label{Ombis}
\Omega_m = \frac{\Omega_{m0} H_0^2}{a^3 H^2}\,,
\end{equation}
where $\Omega_{m0}\equiv\rho_{m0}/(3M_p^2H_0^2)$.  Combining  
relation \eqref{Ombis} with the Friedmann equation \eqref{eqform}, we are 
finally led to
\begin{equation}
    \label{1overHa}
    \frac{1}{Ha}=\frac{1}{H_0}\sqrt{\frac{a\left(1-\Omega_{\mathrm{DE}}\right)}{\Omega_{m0}}}\,.
\end{equation}

Now, by using Eqs.~\eqref{IRc} and  \eqref{BHDE}, together with the definition 
\eqref{frde} of \( \Omega_{\mathrm{DE}} \), we obtain
\begin{equation}
    \int_a^\infty \frac{da'}{H a'^2} = \frac{1}{a} \left( \frac{C}{3 M_p^2 H^2 \Omega_{\mathrm{DE}}} \right)^{\frac{1}{2 - \Delta}}\,,
\end{equation}
which can be simplified by introducing the variable \( x \equiv \log a \), yielding
\begin{equation}
     \int_x^\infty \frac{dx'}{Ha}=\frac{1}{a}  \left( \frac{C}{3 M_p^2 H^2 
\Omega_{\mathrm{DE}}} \right)^{\frac{1}{2 - \Delta}}\,.
\end{equation}
Upon substituting Eq.~\eqref{1overHa}, we acquire
{\small{
\begin{equation}
    \frac{1}{H_0\sqrt{\Omega_{m0}}}\int_x^\infty\sqrt{a\left(1-\Omega_{\mathrm{DE}}\right)}\hspace{0.6mm} dx'=\frac{1}{a} \left( \frac{C}{3 M_p^2 H^2 \Omega_{\mathrm{DE}}} \right)^{\frac{1}{2 - \Delta}}\,.
\end{equation}}}
Differentiating this equation with respect to $x$, we get
\begin{eqnarray}
\nonumber
    \frac{\Omega'_{\mathrm{DE}}} 
{\Omega_{\mathrm{DE}}\left(1-\Omega_{\mathrm{DE}}\right)}&=&1+\Delta + 
Q\left(1-\Omega_{\mathrm{DE}}\right)^{\frac{\Delta}{2\left(\Delta-2\right)}}\\[
2mm]
    &&\times\,\Omega_{\mathrm{DE}}^{\frac{1}{2-\Delta}}\,e^{\frac{3\Delta}{2\left(\Delta-2\right)}x}
    \,,
    \label{Ompr}
\end{eqnarray}
where the prime symbol denotes derivative with respect to \( x \), and we have 
defined the dimensionless parameter 
\begin{equation}
    Q\equiv\left(2-\Delta\right)\left(\frac{C}{3M_p^2}\right)^{\frac{1} 
{\Delta-2}}\left(H_0^2\hspace{0.5mm}\Omega_{m0}\right)^{\frac{\Delta}{
2\left(2-\Delta\right)}}\,.
\end{equation}

The differential equation (\ref{Ompr}) governs the dynamics of Barrow HDE 
within a spatially flat FRW Universe
filled with pressureless matter. 
When the Barrow exponent is set to \( \Delta = 0 \), the framework reduces to the standard HDE model \cite{Li:2004rb}. Indeed, in this limiting case, the evolution equation simplifies to
\begin{equation}
\Omega'_{\mathrm{DE}} \big|_{\Delta = 0} = \Omega_{\mathrm{DE}} \left(1 - \Omega_{\mathrm{DE}}\right) \left( 1 + 2M_p\sqrt{\frac{3\Omega_{DE}}{C}} \right)\,,
\end{equation} 
which admits an implicit analytic solution \cite{Li:2004rb}. Nevertheless, in the general case where the Barrow exponent \( \Delta \) is nonzero, Eq. \eqref{Ompr} exhibits explicit dependence on $x$, and thus requires numerical treatment to obtain the evolution of the dark energy density parameter \cite{Saridakis:2020zol}.

Based on the above formalism, one can further compute the equation-of-state 
(EoS) parameter for Barrow HDE , defined as \( w_{\mathrm{DE}} \equiv 
p_{\mathrm{DE}} / \rho_{\mathrm{DE}} \). To this end, we consider the time 
derivative of Eq. \eqref{BHDE}, which yields
\begin{equation}
\dot{\rho}_{\mathrm{DE}} = \left(\Delta - 2\right) C R_h^{\Delta - 3} \dot{R}_h\,,
\end{equation}
where \( \dot{R}_h \) is obtained using the definition \eqref{IRc}, leading to $\dot{R}_h = H R_h - 1$. With the additional use of $R_h=\left(\dfrac{\rho_{\mathrm{DE}}}{C}\right)^{\frac{1}{\Delta-2}}$, the continuity equation \eqref{contDE} becomes
\begin{equation}
w_{\mathrm{DE}}=-\frac{1}{3}\left[1+\Delta + Q\left(1-\Omega_{\mathrm{DE}} 
\right)^{\frac{\Delta}{2\left(\Delta-2\right)}}
\Omega_{\mathrm{DE}}^{\frac{1}{2-\Delta}}\,e^{\frac{3\Delta}{2\left(\Delta-2\right)}x}\right]
    \,, 
    \label{wde}
\end{equation}
where we have resorted to Eq. \eqref{1overHa} and have rewritten \( 
\rho_{\mathrm{DE}} \) in terms of   \( \Omega_{\mathrm{DE}} \) using 
 \eqref{frde}.
Hence, the dynamics of \( w_{\mathrm{DE}} \) can be determined, provided that 
the evolution of \( \Omega_{\mathrm{DE}} \) is obtained from Eq. \eqref{Ompr}. 
Once again, we point out that  the $\Delta=0$  limit correctly reproduces the 
standard HDE behavior, yielding
\begin{equation}
    w_{\mathrm{DE}}\big|_{\Delta = 0} =-\frac{1}{3}\left(1+2M_p\sqrt{\frac{3\Omega_{\mathrm{DE}}}{C}}\right).
\end{equation}

Lastly, as we mentioned above, Tsallis HDE can be obtained from the above 
expressions under the identification 
\begin{equation}
\delta\rightarrow1+\frac{\Delta}{2}.
\end{equation}

\section{Observational Constraints}
\label{Obs}

In this section we use observational datasets in order to extract constraints 
on the parameters of Barrow and Tsallis holographic dark energy. Let us first 
  describe the data that we use.

\begin{itemize}

\item Observational Hubble Data (OHD): This data set includes 31 direct
measurements of the Hubble parameter from passive elliptic galaxies, known
as cosmic chronometers.\ The measurements for redshifts in the range $%
0.09\leq z\leq 1.965~$as summarized in \cite{cc1}.

\item Pantheon+ (SN/SN$_{0}$): This set includes 1701 light curves of 1550
spectroscopically confirmed supernova events within the range $%
10^{-3}<z<2.27~$\cite{pan}. The data provide the distance modulus $\mu
^{obs}~$at~observed redshifts~$z$. We consider the Pantheon+ data with the
Supernova $H_0$ for the Equation of State of Dark energy Cepheid host distances
calibration (SN$_{0}$) and without the Cepheid calibration.

\item Baryonic acoustic oscillations (BAO): These data are provided by the 
the DESI 2025 Collaboration \cite{DESI:2025zgx,DESI:2025zpo,DESI:2025fii}.
\end{itemize}

For the analysis, we employ COBAYA \cite{cob1,cob2}, with a custom theory,
alongside Markov Chain Monte Carlo (MCMC). We consider the free parameters to 
be the current energy density of
the dark matter, $\Omega _{m0}$, the Hubble constant $H_{0}$, the 
exponent  $\Delta $ and the constant $Q$, as well as the $r_{drag}$ which 
refers to the maximum distance sound waves could travel in the early Universe 
before the drag
epoch.   Moreover, for the initial condition we consider  $\Omega
_{DE}\left( z\rightarrow 0\right) =1-\Omega _{m0}$.

We perform our analysis for different datasets, as well as for various 
combinations, in particular
   SN+BAO and  SN+OHD+BAO, ones. 
We consider the following priors: $H_{0}\in \left[ 65,80\right]$ (in units of \( \mathrm{km \cdot s^{-1} \cdot Mpc^{-1}} \)) and 
$\Omega
_{m0}\in \left[ 0.2,0.4\right] $. Additionally, concerning the $\Delta$ value 
we impose 
$\Delta \in \lbrack 0,1)$, while for the constant $Q$ we use  $Q\in \left[
-2,5\right] $.

\begin{figure}[!]
\centering\includegraphics[width=0.5\textwidth]{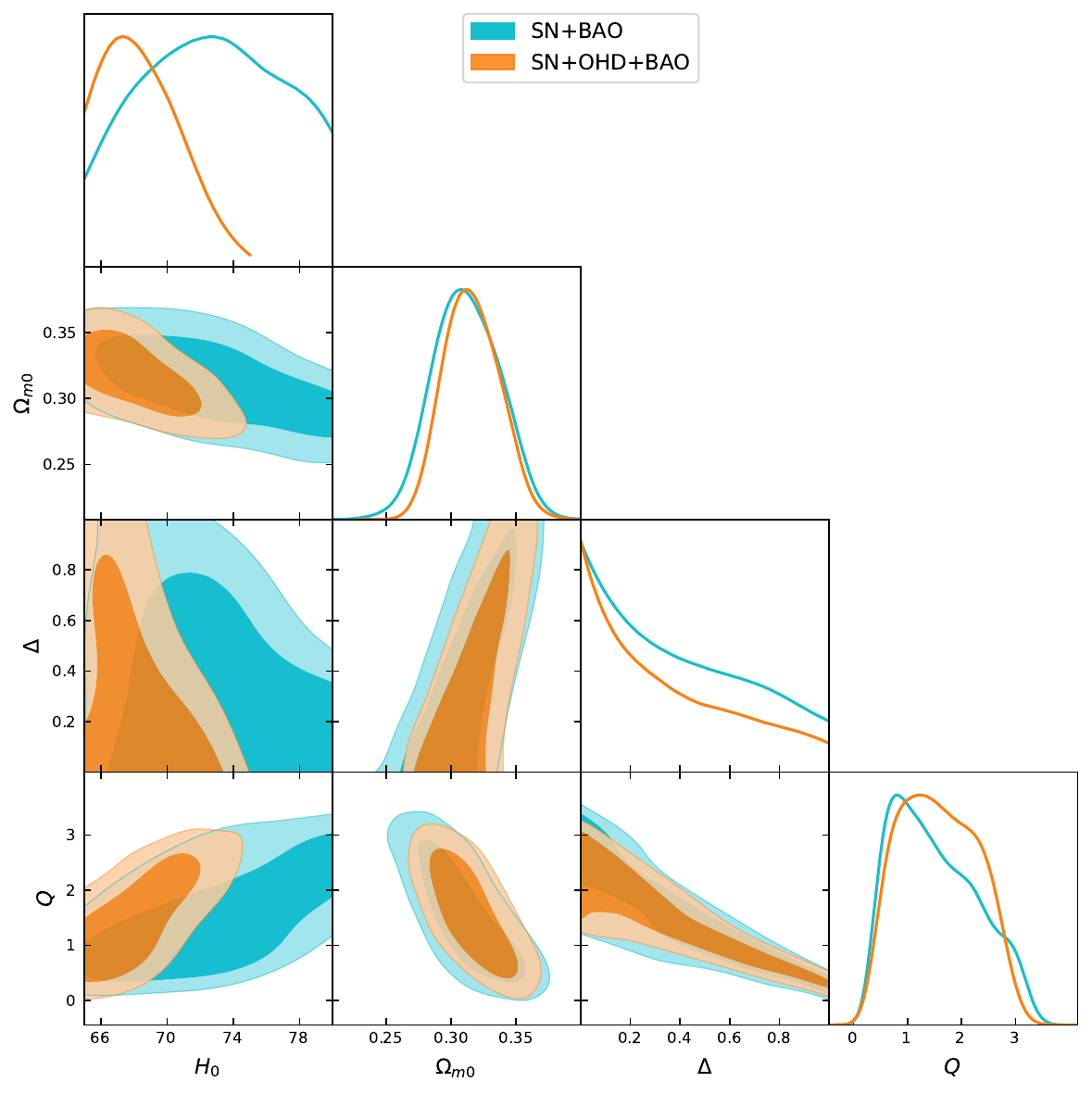}
\caption{{
\it{Likelihood contours for the model parameters of Barrow and Tsallis 
holographic dark energy, for the datasets     SN+BAO and  SN+OHD+BAO. The 
Tsallis exponent $\delta$ is obtained from the Barrow exponent $\Delta$, under 
the 
identification $\delta\rightarrow1+\Delta/2$. }}}%
\label{datafig}%
\end{figure}

The best fit parameters are displayed in Talbe \ref{data1}.
Furthermore, 
 in Fig. \ref{datafig} we draw the iso-likelihood contours 
for the model parameters.
Concerning the value of $\Omega_{m0}$ we observe that we obtain similar results 
with $\Lambda$CDM paradigm. Additionally, concerning the $H_0$ value, we see 
that for    SN+BAO datasets it has the tendency to larger values 
($72.7_{-3.9}^{+3.9}$ ) comparing to  $\Lambda$CDM scenario, however, when the 
full   SN+OHD+BAO datasets are considered it obtains values similar to the 
latter ($68.6_{-3.3}^{+1.3}$), and thus the $H_0$ tension cannot be alleviated
\cite{DiValentino:2025sru}.

 Concerning the exponent $\Delta$ (or similarly $\delta$ for Tsallis cosmology) 
we see that the standard value $\Delta=0$ (and $\delta=1$) lies at the center 
of the contour plots, however the contours spread to positive values (and to 
$\delta>1$ values), similarly to what was found in  
\cite{Anagnostopoulos:2020ctz}. We mention that this is in contrast to the 
results of \cite{Luciano:2025hjn}, in which Barrow and Tsallis exponents were 
confronted with the data, however not in holographic dark energy application but 
in the radically different framework of modified cosmology in the framework of 
gravity-thermodynamics conjecture. In particular, while in 
\cite{Luciano:2025hjn} it was found that $\Delta$ has a tendency to negative 
values, in the present analysis we find a tendency to positive values (in both 
scenarios the standard value $\Delta=0$ lies within the 1$\sigma$ region). 
Actually, this is the reason that the present HDE scenaria cannot alleviate the 
Hubble tension, while in  modified cosmology through gravity-thermodynamics 
conjecture with Barrow and Tsallis entropy it is know that the tension can be 
alleviated \cite{Basilakos:2023kvk}.

\begin{table}[!] \centering%
\caption{Cosmological parameters of Barrow and Tsallis holographic dark energy. 
The Tsallis exponent $\delta$ is obtained from the Barrow exponent $\Delta$, 
under the 
identification $\delta\rightarrow1+\Delta/2$. }%
\resizebox{\columnwidth}{!}{
\begin{tabular}{cccccc}
\hline\hline
\textbf{Barrow Entropy} & $\mathbf{H}${$_{0}$} & $\mathbf{\Omega }_{m0}$ & $%
\mathbf{\Delta }$ & $Q$ & $\mathbf{\chi }_{\min }^{2}$ \\ \hline
\textbf{SN+BAO} & $72.7_{-3.9}^{+3.9}$ & $0.312_{-0.026}^{+0.026}$ & $<0.542$
& $1.51_{-1.10}^{+0.55}$ & $1420.5$ \\ 
\textbf{SN+OHD+BAO} & $68.6_{-3.3}^{+1.3}$ & $0.316_{-0.023}^{+0.020}$ & $%
<0.471$ & $1.57_{-0.88}^{+0.68}$ & $1435.5$ \\ \hline\hline
\end{tabular}%
}
\label{data1}  
\end{table}%

In order to evaluate  the fitting efficiency and compare the behavior of 
Barrow and Tsallis HDE  with      $\Lambda $CDM concordance model, we first fit 
the latter with the 
same datasets, and then we apply  the Akaike Information Criterion 
\cite{AIC} (AIC). This criterion is used to compare the fitting performance of models with different numbers of parameters, which is necessary in the present analysis since the Barrow and Tsallis HDE models include additional parameters compared to the \(\Lambda\)CDM scenario.

\begin{table}[ht] \centering%
\caption{Comparison of   Barrow and Tsallis holographic dark energy  with the 
$\Lambda$CDM scenario.}%
\resizebox{\columnwidth}{!}{
\begin{tabular}{ccccc}
\hline\hline
\textbf{Barrow Entropy} & $\mathbf{\chi }_{\min }^{2}-\mathbf{\chi }%
_{\Lambda \min }^{2}$ & $\mathbf{AIC}-\mathbf{AIC}_{\Lambda }$ & $\mathbf{%
\chi }_{\min }^{2}-\mathbf{\chi }_{B\min }^{2}$ & $\mathbf{AIC}-\mathbf{AIC}%
_{B}$ \\ \hline
\textbf{SN+BAO} & $+0.7$ & $+4.7$ & $+1.1$ & $+3.1$ \\ 
\textbf{SN+OHD+BAO} & $+0.2$ & $+4.2$ & $+0.7$ & $+2.7$ \\ \hline\hline
\end{tabular}%
}
\label{data2}%
\end{table}%

In Table \ref{data2}, we report the differences in $\mathbf{\chi}^2_{\min}$ and $\mathbf{AIC}$ for the Barrow–Tsallis HDE model, evaluated relative to the $\Lambda$CDM scenario (denoted by the subscript $\Lambda$) and the Barrow–Tsallis cosmology (denoted by $B$) \cite{Luciano:2025hjn}.
As observed, although the Barrow and Tsallis HDE scenarios are consistent with the data, they do not exhibit improved performance compared to the $\Lambda$CDM paradigm, nor with respect to the Barrow–Tsallis cosmology.
Nevertheless, as we mentioned above, this 
is not a result against Barrow and Tsallis entropy themselves, rather it is a 
disadvantage of their application within holographic dark energy framework, 
since in other frameworks, such as the gravity-thermodynamics one, they lead to 
viable phenomenology, statistically equivalent with $\Lambda $CDM scenario 
\cite{Luciano:2025hjn}, 
being able to alleviate the $H_0$ tension too \cite{Basilakos:2023kvk}.

\section{Conclusions}
\label{Conc}

Holographic Dark Energy (HDE) is a widely studied model based on  the 
holographic principle of quantum gravity, which postulates that the dark energy 
density is inversely proportional to the square of an infrared (IR) cutoff 
scale, usually taken to be the future event horizon. In order to take into 
account  possible deviations from standard thermodynamics in the 
high-energy or quantum gravity regimes, various extensions of HDE have been 
proposed using generalized entropies.  Tsallis entropy introduces a 
non-additive, power-law correction to the Boltzmann-Gibbs entropy, controlled by 
a parameter $\delta$, which quantifies the degree of non-extensivity, with 
$\delta=1$ corresponding to the Bekenstein-Hawking 
entropy. On the 
other hand,   Barrow entropy introduces quantum gravitational corrections 
arising from a fractal structure of spacetime geometry, parametrized by the 
Barrow exponent $\Delta$, with $\Delta=0$ recovering the Bekenstein-Hawking 
entropy and $\Delta=1$ corresponding to maximal quantum deformation. When 
these modified entropy forms are applied in the HDE framework, one obtains 
Tsallis ad Barrow extended HDE scenarios, with richer phenomenology. 
Interestingly enough, although Tsallis and Barrow entropies have completely 
different theoretical origins, at the level of cosmological equations they 
coincide, under the identification  $\delta\rightarrow1+\Delta/2$.

In this work  we have employed the most recent Baryon Acoustic Oscillation 
(BAO) measurements from the DESI DR2 dataset, alongside data from  Supernova 
Type Ia (SNIa) and Cosmic  Chronometers (CC) observations, to constrain the 
parameter space of both Tsallis and Barrow HDE models.
We found that the DESI data place 
tight bounds on the Barrow exponent, with the best-fit value lying close to 
the standard value $\Delta=0$, however with the bulk of the contour extending 
towards   positive values. This is in contrast with application of Barrow 
entropy within the different framework of gravity-thermodynamics conjecture \cite{Luciano:2025hjn}, 
where it was found that negative values were favoured, and it additionally 
offers an 
explanation why the obtained  $H_0$ values in the present analysis are close to 
those of $\Lambda$CDM cosmology and hence the $H_0$ tension cannot be 
alleviated, while in the gravity-thermodynamics framework it can. 
Finally, in order to examine the statistical efficiency of the fittings, we 
applied the Akaike Information Criterion. As we saw, although Barrow and 
Tsallis holographic dark energy are in agreement with the data, they cannot be 
favoured in comparison to  $\Lambda$CDM paradigm. 

It would be interesting to confront Barrow and Tsallis holographic dark energy 
with the data at perturbative, structure-growth,  level, using 
Cosmic Microwave Background (CMB) temperature and polarization, weak lensing, 
and $S_8$ observations, since such a confrontation could improve their 
behavior. Such an holistic analysis could provide more subtle information on 
whether extended entropies should be used in holographic dark energy framework, 
or within the radically different gravity-thermodynamics conjecture. This 
investigation will be performed in a future project.

\acknowledgments 
The research of GGL is supported by the postdoctoral fellowship program of the 
University of Lleida. GGL and ENS gratefully acknowledge  the contribution of 
the LISA 
Cosmology Working Group (CosWG), as well as support from the COST Actions 
CA21136 - \textit{Addressing observational tensions in cosmology with 
systematics and fundamental physics (CosmoVerse)} - CA23130, \textit{Bridging 
high and low energies in search of quantum gravity (BridgeQG)} and CA21106 -  
\textit{COSMIC WISPers in the Dark Universe: Theory, astrophysics and 
experiments (CosmicWISPers)}. AP thanks the support of VRIDT through 
Resoluci\'{o}n VRIDT No. 096/2022 and Resoluci\'{o}n VRIDT No. 098/2022. AP was 
Financially supported by
FONDECYT 1240514 ETAPA 2025.

\bibliography{Bib}

\end{document}